# Integrating Public Input and Technical Expertise for Effective Cybersecurity Policy Formulation

Hlekane Ngobeni[1] and Mike Wa Nkongolo[2]

**Abstract.** The ever-evolving digital transformation and increased use of technology comes with increased cyber vulnerabilities, which compromise national security. Cyber-threats become more sophisticated as the technology advances. This emphasises the need for strong risk mitigation strategies. To define strong and robust cybersecurity, policies requires an integrated approach of balancing technical expertise with public input. This paper aims to explore strategies used to balance technical expertise and public input to develop effective and robust cybersecurity policies. It also studied how the effective integration of technical expertise with public input is critical to developing effective strategies and resilient cybersecurity frameworks that strengthens national security. A lack of a holistic approach and collaborative efforts to cybersecurity can hinder the effectiveness of cybersecurity policies. This paper followed a systematic literature review with bibliometric analysis using the PRISMA methodology to explore how technical expertise and public input can be integrated to guide cybersecurity policy making. The thematic analysis identified five important themes in developing effective cybersecurity policies, these key themes are: Multi-Stakeholder Involvement and Human Centric Approaches (MSI & HCA), Governance and Policy Frameworks (GPF), Technical Infrastructure (TI), Evaluation and Compliance (EC), and Legal Rights and Sovereignty (LRS). The synthesis shows that there is no adequate exploration of collaborative efforts which undermines the effectiveness of the cybersecurity policies. The findings suggest that inclusive, flexible governance strategies that integrate public input at every stage are necessary for future cybersecurity policy research and practice, which must shift away from a primarily technical and legal perspective.

## 1 Introduction

Cybersecurity can compromise national security, especially in the age of increased reliance on digital technologies. The rapid technological advancement introduces new vulnerabilities compromising security of sensitive information, public safety, infrastructure and potential economic disruptions (AlDaajeh & Alrabaee, 2024). To remain active participants in the digital economy, states, institutions and the public must adapt to the ever-evolving digital transformation, emphasising the need for strong risk mitigation strategies (Alqurashi & Ahmad, 2024; Ojo, 2021). Defining robust cybersecurity policies to effectively mitigate cyberthreats requires

[1] Department of Informatics, University of Pretoria, South-Africa, hlekane.ngobeni@up.ac.za
[2] Department of Informatics, University of Pretoria, South-Africa, mike.wankongolo@up.ac.za

collaborative efforts, a balance of technical expertise and public input (Salomon, 2022).

However, balancing the integration of public input and technical expertise in the development of cybersecurity policies processes remains a challenge, policy makers often overlook public participation and prioritise technical expertise (Khan, Shiwakoti, Diro, Molla, Gondal & Warren, 2024).

The growing usage of Information Communication Technologies (ICT) such as smartphones, cloud computing, and the Internet of Things (IoTs) has created a new ecosystem of increased human-technology interaction, allowing cybercriminals to exploit valuable assets, critical infrastructures and data breaches; thus, effective cybersecurity policies are needed to respond to the evolving cyber threats (Mishra, Alzoubi, Anwar & Gill, 2022).

Striking a balance between technical and public input is imperative because security is not only a technical issue but a societal issue. Therefore, policy makers leverage the power of collaboration to build trust and transparency to help develop effective cybersecurity regulatory frameworks that will address societal concerns and complex cyber risks (Joubert & Nkongolo, 2025; Khan et al., 2024).

Studies highlight the need for holistic policies which require regulations and frameworks to address cyber risks. For instance, European nations have implemented regulations such as the General Data Protection Regulations (GDPR) to protect sensitive information, identity theft and mitigating cyberattacks (Mishra et al., 2022). Cybersecurity is not a standalone issue, it requires a collaborative approach across multidisciplinary fields (Szczepaniuk & Szczepaniuk, 2022).

This study proposes that the integration of public input and technical expertise can benefit the development of robust cybersecurity policies. The limited participation of diverse actors undermines the development of effective and resilient cybersecurity regulations; and existing research does not directly discuss the inclusion of the general public as part of the broader strategy.

To address this challenge, this paper aims to explore strategies that can be employed to balance technical expertise and public input in cybersecurity policy development by answering the research question: *What strategies can be used to integrate technical expertise and public participation for developing effective and practical cybersecurity policies?*

This question is addressed by conducting a Systematic Literature Review (SLR) using a PRISMA methodology. Data was collected from these academic databases; *Science Direct, Emerald Insight and Scopus*, and the following keywords were used to search across all three databases: *"Cybersecurity Frameworks" OR "Cybersecurity policies" AND "National Security" AND "Public participation" OR "Public Engagement" AND "Federal Agencies"*.

A consistent inclusion and exclusion criteria were applied. Only journal articles published in the year 2020 to 2025 were included. The extracted metadata was sorted using EndNote and Excel to remove duplicates, articles not written in English and articles that did not align to the research question.

The remainder of the study is organised as follows: Section 2 presents a review of literature on cybersecurity policy frameworks and multistakeholder participation.

Section 3 provides the research methodology, and Section 4 provides the findings and thematic analysis. Section 5 presents the discussion and synthesis of these findings. The concluding remarks are presented in Section 6, and finally, Section 7 outlines recommendations for future research.

## 2 LITERATURE BACKGROUND

As the world becomes increasingly interconnected, the human-technology interaction ecosystem is vulnerable to cyber threats and more prone to cyber-attacks, making cybersecurity an important component in cyberspace (Mishra et al., 2022). Cybercriminals use the cyberspace to target valuable assets, critical infrastructure, data breaches, and other critical operations, causing disruptions and instability (Büyüközkan & Güler, 2025; Mishra et al., 2022). These challenges require proactive measures to mitigate cyber risks and safeguard the digital ecosystem (Mishra et al., 2022).

Governments worldwide have realised the criticality of cybersecurity as a strategy to counter cyber threats by adapting frameworks such as the National Cybersecurity Strategic Plan (NCSP) in pursuit of developing a robust and coordinated cyber ecosystem (AlDaajeh & Alrabaee, 2024).

According to the International Telecommunication Union (ITU, 2024), 132 countries have developed National Cybersecurity Strategies (NCS); however, stakeholder engagement is not consistent throughout the policy development cycle.

Governments should understand the importance of comprehensive regulatory frameworks for reliable and responsible governance in cyberspace (Praditya, Maarif, Ali, Saragih, Duarte, Suprapto & Nugroho, 2023).

For example, Indonesia established the National Action Plan (NAP) and Security Knowledge Sharing Network (SKSN) frameworks to safeguard their cyberspace and align with national cybersecurity objectives (Praditya et al., 2023).

These frameworks prioritise transparency, accountability, responsibility, independence and fairness to build public trust ensuring compliance and risk management through cybersecurity skills development (Praditya et al., 2023).

Also, the European Union (EU) has a skills framework that prioritise behavioural cybersecurity, adaptive training, and tackling new issues like Artificial Intelligence (AI) security and small and medium-sized business (SME) requirements (Almeida, 2025). The analysis highlights places where different frameworks diverge and converge, pointing up gaps that call for more flexible and inclusive methods (Almeida, 2025).

Cybersecurity goes beyond technical challenges, for instance, the United States (US) incorporated capacity building and technical assistance as a mechanism to effectively fight against cybercrime (Büyüközkan & Güler, 2025). Frameworks such as the "Cybersecurity: A Guide to Developing National Cybersecurity Strategies," published by the Organisation of American States (OAS) offer best practices in implementing cybersecurity strategies, allowing the OAS member countries effective and adaptive strategies to counter cyber threats and strengthen their cybersecurity resilience (AlDaajeh & Alrabaee, 2024).

Joubert and Nkongolo (2025) emphasise the importance of public perceptions in shaping the legitimacy and effectiveness of cybersecurity policy, arguing that governments should integrate perspectives from both the public and technical experts to enable the design of holistic and inclusive policies. Involving diverse stakeholders will enable governments to formulate robust and inclusive policies (Joubert & Nkongolo, 2025). As such, this paper examines strategies used to balance technical expertise with public input in the development of cybersecurity frameworks.

## 3 RESEARCH METHOD

The research used a Systematic Literature Review (SLR) as the primary research strategy to evaluate and interpret existing literature relevant to the cybersecurity policy making processes, identify gaps and propose strategies to integrate public input and technical expertise in the development of these policies (Khan et al., 2021; Büyükozkan & Güler, 2025). It follows the guidelines of the Preferred Reporting Items for Systematic reviews and Meta-Analyses (PRISMA) to report the process of this systematic review, which involves formulating research questions, search strategies to gather relevant data, and defining the inclusion and exclusion criteria to ensure that relevant sources are retrieved (Büyüközkan & Güler, 2025).

### 3.1 DATA SOURCES AND SEARCH TERMS

The search string was consistently applied on all three databases, with a set of inclusion and exclusion criteria to ensure that relevant and reliable sources are retrieved to answer the research question.

### 3.2 SELECTION CRITERIA

To determine eligibility of studies, the sources were systematically identified, selected and evaluated using a uniform search strategy and applied consistent inclusion and exclusion criteria. The articles selected met the following inclusion criteria: *Studies published in English; Journal articles and research articles; papers that are not older than five years, published between 2020 and 2025*. The articles that did not meet the criteria were excluded as follows: *Duplicates; articles that were inaccessible; pre-prints, and non-academic articles*.

### 3.3 PRISMA FLOWCHART

Screening and evaluation of the studies followed the PRISMA methodology, beginning with 2,270 identified records. After applying the selection criterion and filtering by publication date, 597 records were removed, leaving 1,673. A further filter based on publication type reduced this number to 859. Following duplicate removal, 856 unique records remained. Subsequent screening using the predefined inclusion and exclusion criteria along with title and abstract evaluation resulted in the exclusion of 828 articles. Ultimately, 28 studies were deemed eligible and included in the final synthesis (Figure 1).

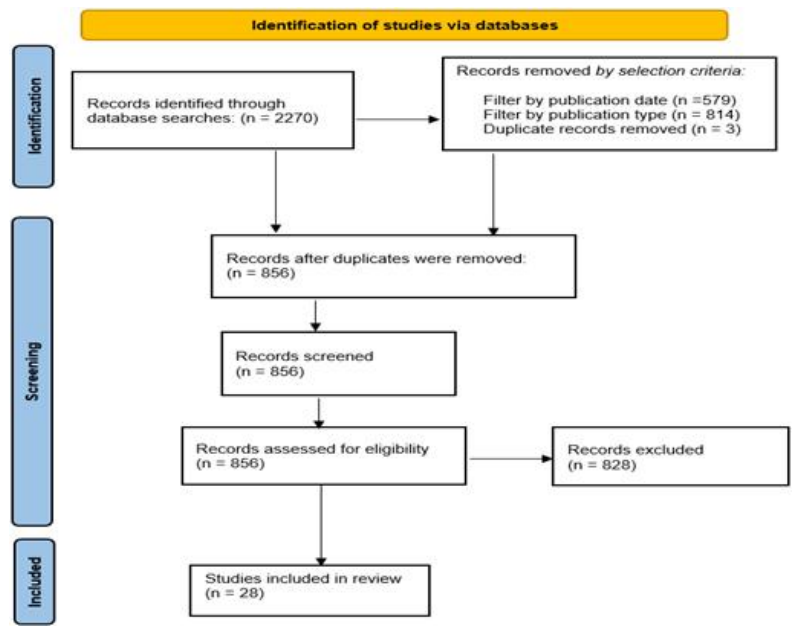

**Fig. 1.** PRISMA flow chart.

### 3.4 DATA EXTRACTION AND ANALYSIS

In this paper, the data extraction process was performed to sort, screen and analyse the selected articles, and the acquired bibliographic data was extracted into an excel spreadsheet to structure and organise the retrieved data for further analysis. A thematic analysis of the results was conducted to identify and present recurring patterns and themes in the extracted data (Sovacool, Iskandarova & Hall, 2023). It allows for the mapping of the datasets, to derive relationships between one study and another, to determine the link between the data and pattern recognitions by identifying themes, sub-themes and the keywords.

According to Sovacool *et al.,* (2023: p3) "thematic analysis extracts meaning from data and encompasses the pinpointing, sharpening, recording, and/or evaluation of recurring themes." This SLR adopts a qualitative analytic method to identify patterns and themes within the dataset and to guide the discussions of the results (Ali, Abdelbaki, Shrestha, Elbasi, Alryalat & Dwivedi, 2023). A bibliometric mapping approach and a thematic analysis were used to visualise the relationships of the extracted data. An online tool, WordArt was used to indicate keyword frequency. The bigger text words are the frequently mentioned keywords (Figure 2). A keyword frequency word cloud highlighting the most common keywords in the literature is shown in Figure 2. The categories *"governance or legislation," "policy," "stakeholders or perception," and "security or data protection"* collectively highlight the importance of integrating technical expertise with public input in effective cybersecurity policymaking. While governance, legislation, and security highlight the importance of technological and legal frameworks, the regular appearance of stakeholders, pereception, and public engagement indicates growing recognition of participatory approaches.

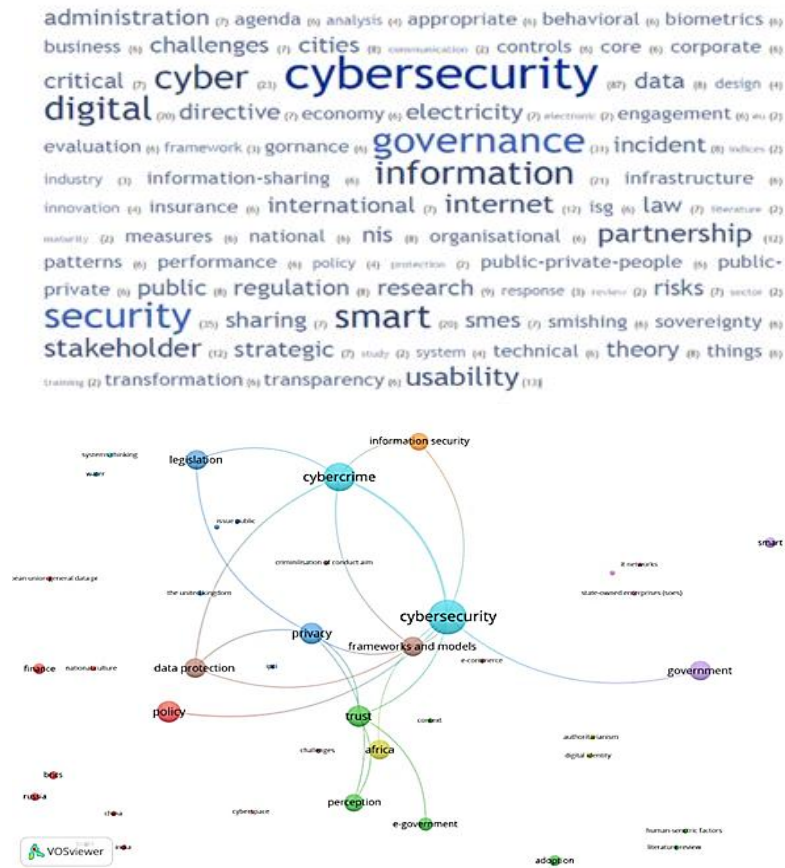

**Fig. 2.** Keywords frequency word cloud visualisation.

Figure 3 shows the distribution of the final 28 articles over the five-year period 2020 to 2025 to give context for the literature's recent focus. This presentation gives context to the extent to which the research is being conducted in cybersecurity. The analysis considers the research strategies, country of study, data analysis techniques, frameworks, keywords used, the recurring patterns, and relationships of key themes and the sub-themes. This method is adapted to present a picture of the relationships between studies, a network of shared perspectives and the gaps in the existing body of knowledge (Sovacool *et al.*, 2023).

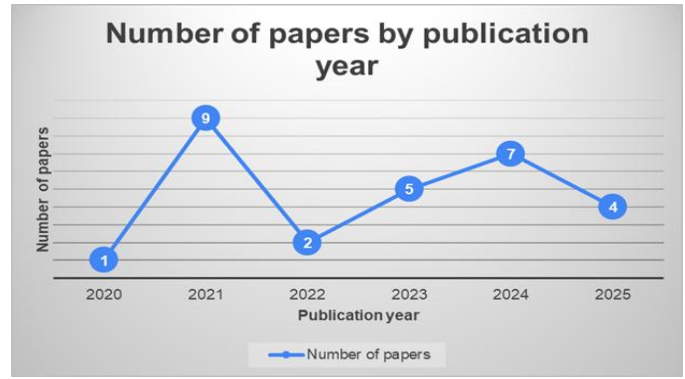

**Fig. 3.** Number of papers by publication year.

## 4  FINDINGS

The findings indicate a growing interest in understanding the cybersecurity landscape and the rising need to build robust cybersecurity frameworks. The growth in publications and more research across multiple regions shows that it is a global concern (AlDaajeh & Alrabaee, 2024). The thematic analysis of the identified articles highlights the importance of collective efforts and collaboration through private-public partnerships, which will support the development of relevant and adaptive cybersecurity policies (Naqvi, Clarke & Porras, 2021). The geographic analysis of the 28 articles shows that a significant body of work comes from the EU (37%) indicating the region's commitment to coordinated efforts to cybersecurity policy.

Multi-country studies make up about 11% while Asia and Latin America each contribute about 16%. Only 5% of research comes from the combination of the US and EU with 10% from the United Kingdom (UK). With little representation from North America and other non-EU regions, this distribution shows a regional imbalance (Figure 4). The focus of study in the EU indicates that further research is needed to examine cybersecurity policy making across a wider range of country settings and governance structures (Figure 4). In another analysis, approximately 35% of the studies focused on developing economies. Notable examples include research in Africa, specifically Malawi (Mtegha et al., 2025), Nigeria (Onatuyeh et al., 2025), and Ethiopia (Ajebo & Solomon, 2024). In these studies, 60% employed qualitative methods such as interviews, focus groups, or case study designs, while 30% used survey-based approaches to gather data on participant knowledge, engagement, and trust levels (Figure 4). In larger capacity-building or educational studies, mixed-methods approaches were commonly applied (Figure 4).

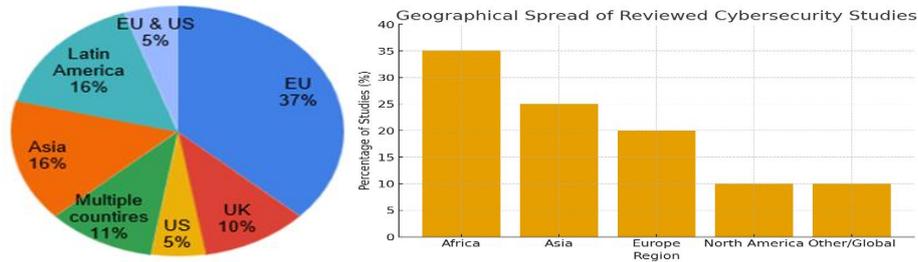

**Fig. 4.** Geographical distribution of cybersecurity policy studies.

The thematic mapping presents a summary of the findings (Figure 5), identifying five main themes which are: Multi-Stakeholder Involvement and Human-Centric Approaches (MSI & HC), Governance and Policy Frameworks (GPF), Technical Infrastructure (TI), Evaluation and Compliance (EC), and Legal Rights and Sovereignty (LRS). Collectively these themes indicate the importance of integrating public participation and technical expertise into developing strong and robust and resilient cybersecurity policies. Figure 5 presents the distribution of sub-themes and total publications among the five main themes showing significant differences in the overall research focus.

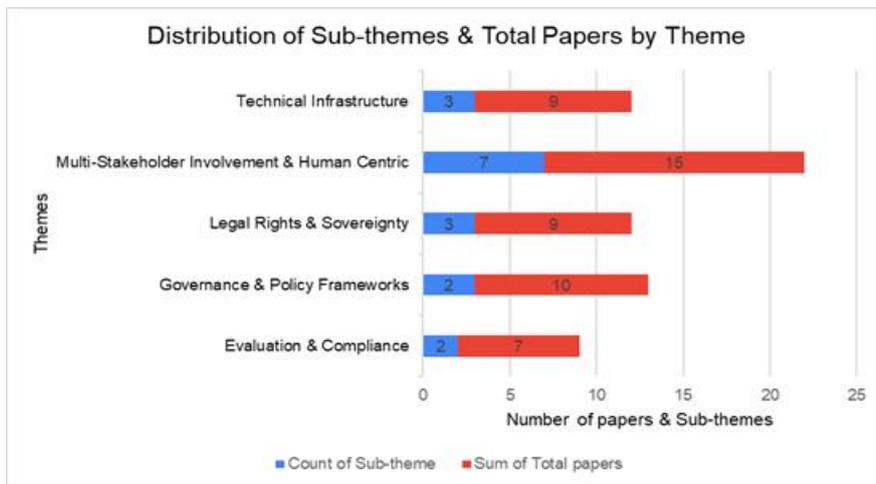

**Fig. 5.** Distribution of sub-themes across main themes and total papers by theme.

The MSI and HC dominate with seven sub-themes and 15 articles. This dominance is a result of an increasing recognition that cybersecurity cannot be solved purely technically and emphasising the importance of collaborative approaches to cybersecurity policies. The TI and LRS, with three sub-themes and nine articles, respectively, highlights an ongoing interest in the technological aspects and contribution of compliance systems to building cybersecurity resilience (Figure 5). The two themes remain relevant as they support the creation of safe infrastructure, regulatory compliance and new tools that improve cyber resilience. The GPF with two themes and 10 articles, also shows the continued focus on the importance of policies and regulatory frameworks to strengthen cybersecurity. EC appears to be the least studied area, with two themes and seven articles (Figure 5). This indicates that more effort is needed in building evaluation processes to assess and enforce compliance. All these findings highlight a changing research environment that is progressively shifting away from traditional, expert-driven methodologies toward approaches that prioritise societal values, adaptability, and inclusiveness in shaping cybersecurity governance.

### 4.1. Multi-Stakeholder Involvement and Human Centric

The development of functional and resilient cybersecurity governance requires active collaboration among multiple stakeholders, that is the government, private sector, civil society, academia and citizens (Naqvi et al., 2021). The human aspect is a critical factor and an overlooked component in cybersecurity design, considering that the users play a critical role in developing functional systems and regulations (Naqvi et al., 2021). To build a cyber-resilient ecosystem, it is crucial to bridge the technological skills gap to enable users to understand cyber-threats and to effectively handle the threats (Khanna & Khanra, 2023). Investing in building capacity through skills development and awareness initiatives will foster a culture of cyber consciousness and resilience (AlDaajeh & Alrabaee, 2024). The EU, for example, has established a cybersecurity framework to improve cybersecurity competency guidelines and to upskill cybersecurity professionals to respond to the fast-evolving industry demands (Almeida, 2025).

### 4.2. Governance and Policy Framework

Developing practical cybersecurity initiatives requires a solid foundation in governance and policy frameworks. They are the cornerstone of national cybersecurity plans that are designed to manage cyber threats and protect digital infrastructure (Khan, Shiwakoti, Stasinopoulos; Chen & Warren, 2025). Good governance with clearly defined stakeholder roles promotes trust, legal and regulatory compliance (AlGhamdi et al., 2020). Smart governance is becoming a crucial component of cybersecurity policymaking focusing on improving governance procedures through data-driven decision-making and the inclusive involvement of civic actors in the formulation and application of policies, going beyond the simple adoption of new technologies (Jurado-Zambrano, Velez-Ocampo & López-Zapata, 2022). The inclusion of well-known information security frameworks such as COBIT, ISO/IEC 27001, and ITIL offers a structured foundation for developing successful security control objectives and enhancing the overall cybersecurity posture. By facilitating the integration of operational security needs with strategic governance objectives, these frameworks help create governance systems that are more flexible and robust (Jurado-Zambrano et al., 2022).

### 4.3. Technical Infrastructure

As Putro, Sensuse & Wibowo (2024) discussed in their paper, smart government requires a strong infrastructure and regulations, such as rules governing digital technologies and the importance of establishing security controls around critical security infrastructure. For example, Indonesia's security framework is used as a guideline to regulate security standards and safeguard critical infrastructure (Putro et al., 2024). A successful digital operation, flow of data and services is dependent on networks and infrastructure which can be prime targets for cyber-attacks, therefore making this area a critical aspect in cybersecurity. Technical infrastructure protection is fundamental and must be embedded within the broader governance framework (Almeida, 2025).

### 4.4. Evaluation and Compliance

Effective cybersecurity governance depends on continuous assessment and compliance procedures to ensure that regulations are implemented in a practical way (Schmitz, Schmid, Harborth & Pape, 2021). However, Schmitz et al., (2021) identified a gap between technical capability and effective evaluation as security control maturity assessments faced issues of accuracy and reliability due to insufficient organisational support. Cybersecurity maturity models play a critical role in bridging technical knowledge with legal and regulatory compliance (Koolen, Wuyts, Joosen & Valcke, 2024). According to Koolen et al., (2024) these maturity models facilitate compliance with the General Data Protection Regulation (GDPR), offering structured, and risk-aligned assessment tools as guidelines to help organisations meet their legal obligations. Additionally, policy learning through incident reporting and feedback loops is important for security control readiness and adaptive governance. Busetti and Scanni (2025) states that it requires effective communication and regulatory enforcement to translate the lessons learned from the incident reports and feedback loops into improved cybersecurity regulatory frameworks.

### 4.5. Legal Rights and Sovereignty

Zieliński (2024) explores the legal landscape by looking at Poland's Act of Combating Abuses in Electronic Communications and the EU's cybersecurity strategy for the digital age. They use Poland's as an example on how they respond to threats, demonstrating how legislation can enforce compliance while engaging public and private sectors in ensuring cybersecurity best practices (Kulesza & Weber, 2021). This emphasises the importance of frameworks that allow participation from multi-stakeholders. Katsikas (2025) also highlights the importance of digital sovereignty, to build trust, transparency, and resilience; nations must strive for security control over their digital infrastructure and cyber operations and thus influences national cybersecurity strategy (Katsikas, 2025; Zieliński, 2024).

# 5  DISCUSSION

The continued reliance on digital technologies in a globalised society has necessitated the need for building resilient and practical cybersecurity policy frameworks to ensure protection of critical infrastructure, sustainable development and national security (AlDaajeh & Alrabaee, 2024; Naqvi *et al.*, 2021). This systematic literature review studies the integration of public input and technical expertise to effectively build robust and resilient cybersecurity policies.

The findings emphasise the perspective that cybersecurity is not only a technical issue, but a societal and governance issue that requires collaborative efforts and shared responsibility to help develop regulatory frameworks that guides the cyber landscape (Naqvi *et al.*, 2021).

Although there is a growing interest in cybersecurity regulatory standards, legal frameworks, and infrastructure protection, there is no adequate exploration of collaborative governance strategies. Limited involvement of key actors, including citizens, SMEs and civil society hinders the effectiveness of cybersecurity policies (Khan *et al.,* 2024; Almeida, 2025).

The thematic analysis puts high emphasis on infrastructure protection, compliance and national security, with less focus on participatory governance and human centric approaches, while sub-themes such as civil society inclusion, citizen engagement and public-private partnerships receive less research attention (Naqvi *et al.,* 2021; Alghami *et al.,* 2020; Schmitz *et al.,* 2021).

This imbalance indicates a research gap in cybersecurity policy strategies where the multi-stakeholder involvement remains overlooked, leading to ineffective cybersecurity strategies.

Multi-stakeholder involvement is important in building successful governance frameworks. Smart governance principles that support transparency and digital responsiveness are necessary for aligning with the cybersecurity strategies (Khan et al, 2025; Jurado-Zambrano et al., 2022), and established frameworks provide structured approach for aligning security controls with strategic goals while maintaining regulatory compliance (Almeida, 2025; Katsikas, 2025; Zeilinski, 2024).

Evaluation and compliance strategies provide continued security control through maturity assessments, to measure compliance with regulatory standards that compliments the protection of technical infrastructure (Koolen *et al.*, 2024; Schmitz *et al.,* 2021). These strategies facilitate compliance with regulation like the GDPR by offering structured, risk-aligned assessment tools as guidelines to help organisations meet their legal obligations and therefore ensuring the protection of critical infrastructure (Koolen *et al.*, 2024).

Incorporating legal rights and sovereignty into cybersecurity policies is crucial in ensuring the protection of citizen's digital rights and national security (Kulesza & Weber, 2021; Zieliński, 2024). Drawing from the thematic analysis, the conceptual model formulated in this paper proposes an integrated framework in which technical infrastructure, legal sovereignty and evaluation strategies serve as the foundation for effective cybersecurity governance, all interconnected through the involvement of various stakeholders (Almeida, 2025). The knowledge gap between technical and non-technical stakeholders is addressed through capacity building, serving as enablers for public involvement (Khanna & Kanrah, 2023). This model emphasises an ongoing feedback loop that supports continuous development and adaptability through iterative policy improvements driven by incident reporting, policy learning and public input (Busetti and Scanni, 2025). According to this synthesis, inclusive and flexible governance strategies that integrate public input at every stage are necessary for future cybersecurity policy research and practice. It shifts from a primarily technical and legal perspective which improves not only trust but also allow for stakeholder collaborations (Figure 6).

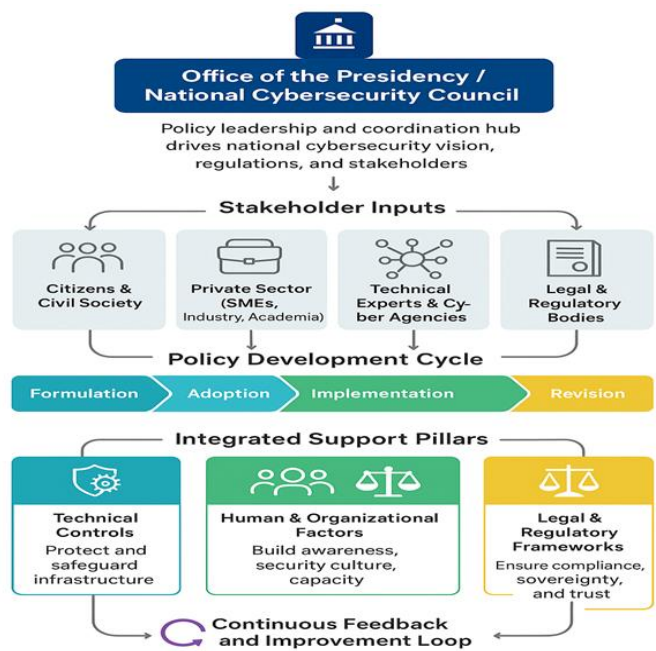

**Fig. 6.** Proposed framework for cybersecurity policy development.

Based on the synthesis, Figure 7 shows a process-oriented, integrated framework for resilience and cybersecurity policy formulation. Putting the Office of the Presidency as the primary coordinating body in charge of stakeholder alignment (Figure 6), regulatory coordination, and national cybersecurity strategy.

To create inclusive cybersecurity policies, the model incorporates a multi-stakeholder approach that acknowledges the collaborative efforts of citizens, the private sector, civil society, technical experts, and legal professionals (Naqvi et al., 2021; Almeida, 2025). The four interconnected phases of the policy-development cycle include formulation, adoption, implementation, and revision which are all informed by ongoing stakeholder participation and directed by strategic objectives (Figure 7).

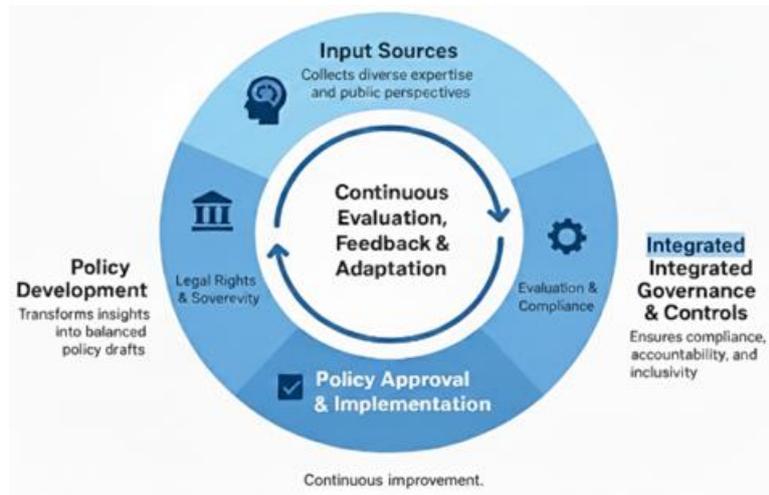

**Fig. 7.** The proposed four-layer cybersecurity framework.

Its supporting pillars are technical controls, human and organisational factors, legal and regulatory frameworks (Figure 7). A continuous feedback and adaptation loop at its core allows for responsiveness to new threats through regular assessment and inclusive public participation (Busetti & Scanni, 2025). This study proposes a four-layer cybersecurity policy framework (Figure 7) that integrates technical expertise, public input, governance, and compliance within an ongoing feedback cycle (Naqvi *et al.,* 2021; Almeida, 2025). The first layer promotes various input sources from experts and the public to ensure inclusion and trust (Schmitz *et al.,* 2021; Khan *et al.,* 2024). The second layer integrates these into policy development, measuring technical feasibility against social needs (Alghami *et al.,* 2020; Katsikas, 2025). The last two layers emphasise the importance of regulation, governance and implementation, layer three improves governance and operational controls to ensure responsibility, compliance, and protection of rights (Kulesza & Weber, 2021; Zieliński, 2024). And to ensure feasibility and practicality of these policies, layer four looks at policy approval and execution (Almeida, 2025; Zeilinski, 2024). This shows that cybersecurity resilience is built on a continuous process of collaboration, learning and development, which is supported by inclusive governance frameworks (Figure 6-7).

## 6    CONCLUSION

This paper has systematically reviewed relevant literature with bibliometric analysis following the PRISMA methodology guidelines to evaluate and interpret existing literature relevant to the cybersecurity policy making processes. It looked at how public input and technical expertise be integrated into cybersecurity policy making to ensure the protection of critical infrastructure, strengthen national security and improve cyber resilience. Thematic analysis of the selected articles identified five key themes that are important in developing effective cybersecurity policies, these include Multi-Stakeholder Involvement and Human-Centric (MSI & HC), Governance and Policy Frameworks (GPF), Technical Infrastructure (TI), Evaluation and Compliance (EC), as well as Legal Rights and Sovereignty (LRS). A significant gap was identified between the importance of multi-stakeholder involvement and incorporating it in cybersecurity policy development.

The synthesis of the findings emphasises that defining strong and robust cybersecurity policies will remain insufficient without balanced and inclusive governance strategies. Highlighting that cybersecurity is not a matter of technical issues only, but it is a societal and governance issue that requires collaborative efforts and shared responsibility to help develop regulatory frameworks that are practical and effective. And thus, the inclusion of diverse stakeholder perspectives ensures that policies are not only technically sound but also socially acceptable and practical, which improves compliance, resilience and public trust.

The conceptual model suggests an integrated framework in which technical infrastructure, legal sovereignty and evaluation strategies serve as the foundation for effective cybersecurity governance policies, all of which are interconnected through stakeholder engagement.

The contribution of this paper is to develop a practical framework that incorporates public input and technical expertise, which serves as a practical guide to policy makers for building policies that are feasible, inclusive and flexible to handle the constantly changing online risks. Its emphasis on collaborative governance and contentious feedback loops promotes policy feasibility, resilience, public trust and compliance.

Therefore, future cybersecurity research must shift from the traditional technical and legal perspectives towards inclusive, flexible governance strategies that integrate public involvement at every stage of policy development. This shift will not only improve trust, but also allow policy makers to develop robust, inclusive and resilient cybersecurity ecosystems.

Future research should prioritise developing and testing practical frameworks that explore multistakeholder involvement in cybersecurity policy making. Policy makers and research must focus on effectively facilitating public involvement, through capacity building and skills development which serves as enablers for public involvement. Moreover, studies should explore methods and metrics for evaluating the effectiveness of inclusive governance practices and their influence on trust, compliance and resilience. The practical application of frameworks will be further enhanced by research that examines their contextual adaptation across various regulatory, socio-political, and technical situations. Finally, exploring the integration of changing cybersecurity threats into participatory policymaking processes would help frameworks remain adaptable and future ready.